\documentclass[modern]{aastex61}

\usepackage{graphicx}
\usepackage{amsmath}
\usepackage{amssymb}
\usepackage{enumitem}

\newcommand\mybar{\kern1pt\rule[-\dp\strutbox]{.8pt}{\baselineskip}\kern1pt}

\setlist[itemize]{noitemsep, topsep=0pt, leftmargin=*}

\shorttitle{Novel Tests of General Relativity}
\shortauthors{Loeb}

\begin{document}

\title{Seven Novel Observational Tests of General Relativity}

\author{Abraham Loeb}
\affiliation{Astronomy Department, Harvard University, 60 Garden
  St., Cambridge, MA 02138, USA}

\begin{abstract}
I study seven novel observational tests of general relativity.  First,
I show that a gravitational wave pulse from a major merger of massive
black holes at the Galactic center induces a permanent increase in the
Earth-Moon separation. For black holes of mass $\sim 10^6M_\odot$, the
shift in the local gravitational potential is comparable to the
Earth-Moon potential. The permanent increase in the Earth-Moon
separation is a fraction of a millimeter, measurable by lunar ranging
for future merger events. Second, I show that General Relativity sets
an absolute upper limit on the energy flux observed from a
cosmological source as a function of its redshift. Detecting a
brighter source in gravitational waves, neutrinos or light, would flag
new physics. Third, I consider the implications of modified inertia at
low accelerations for rockets. An attractive interpretation of
MOdified Newtonian Dynamics (MOND) as an alternative to dark matter,
changes the inertia of matter at accelerations $a\lesssim a_0\approx
1.2\times 10^{-8}~{\rm cm~s^{-2}}$. I show that if inertia is modified
at low accelerations, this suppresses the exponential factor for the
required fuel mass in low acceleration journeys. Rockets operating at
$a\ll a_0$ might allow intergalactic travel with a modest
fuel-to-payload mass ratio.  Fourth, I show that in MOND the amplitude
of the observed dipole of the Cosmic Microwave Background (CMB) can
originate from the primordial fluctuation amplitude on the scale of
the cosmic horizon. Fifth, I show that the tidal gravitational
potential of the Milky-Way galaxy removes fuzzy dark matter from its
satellite dwarf galaxies through quantum-mechanical tunneling.  Sixth,
I show that a charged particle can be accelerated to arbitrarily high
energies by maintaining a permanent resonance with the phase of a
planar gravitational wave propagating along a uniform magnetic
field. The Doppler-shifted cyclotron autoresonance could potentially
result in electromagnetic afterglows near gravitational-wave sources.
Seventh, I show that non-relativistic objects traveling in
intergalactic space at an initial peculiar velocity $v$ will traverse
in the future of the LCDM cosmological model a maximum comoving
distance of $\sim (v/1.7H_0)$, irrespective of travel time, where
$H_0$ is the Hubble constant.  To reach beyond the Virgo cluster of
galaxies, requires an initial peculiar speed $\gtrsim 3\times
10^3~{\rm km~s^{-1}}$, a hundred times faster than the chemical
rockets launched to space so far.

\end{abstract}

\section{Detecting the Memory Effect from a Massive Black Hole Merger at
  the Galactic Center Through Lunar Ranging}

The black hole at the Galactic Center, SgrA*, grows in part through
mergers of black holes of mass $M_{\rm BH} \sim 10^6~M_\odot$
\citep{2011MNRAS.414.1127M,2020ARA&A..58..257G}. Here we calculate
the imprint of such mergers on the Earth-Moon seperation.

A merger between black holes of the above mass results in a
gravitational wave pulse of a characteristic duration,
\begin{equation}
(\Delta t)_{\rm GW} \sim \left({GM_{\rm BH}\over c^3}\right) \sim 5~{\rm s}~.
\label{one}
\end{equation}
The mass equivalent of the radiated energy, $(\Delta M)_{\rm GW}$,
changes the near-Earth gravitational potential by an amount,
\begin{equation}
(\Delta \phi)_{\rm GW} \sim {G(\Delta M)_{\rm GW}\over d_{\rm GC}}=
  0.6\times 10^{-11} c^2 \left[{(\Delta M)_{\rm GW}\over M_{\rm
      BH}}\right]~,
\label{two}
\end{equation}
where $c$ is the speed-of-light, $d_{\rm GC}\approx 8~{\rm kpc}$ is
the distance of the Galactic center from the Sun
\citep{2019ApJ...885..131R}, and typically $(\Delta M)_{\rm
    GW}\lesssim 0.1 M_{\rm BH}$ \citep{2014PhRvD..90j4004H}.

Coincidentally, this shift in gravitational potential as a result of
the energy carried by the pulse happens to be comparable to the
gravitational potential that binds the Moon to Earth,
\begin{equation}
\phi_{\Earth} ={GM_{\Earth} \over d_{\rm Moon}}= 10^{-11} c^2~,
\label{three}
\end{equation}
where $M_\Earth=6\times 10^{27}~{\rm g}$ is the mass of the Earth and
$d_{\rm Moon}\approx 4\times 10^{10}~{\rm cm}$ is the Earth-Moon distance.

The gravitational radiation pulse traverses the Earth-Moon system over
a light-crossing time $(\Delta t)_{\rm cross}\sim (d_{\rm Moon}/c)\sim
1.3~{\rm s}$, during which the gravitational-potential change affects
one of the objects before the other. For $(\Delta \phi)_{\rm
  GW}\lesssim \phi_\Earth$, the temporary weakening of the gravitational
binding between the Earth and the Moon during the passage period
$(\Delta t)_{\rm cross}$ leads to an increase in the Earth-Moon
separation by an amount,
\begin{equation}
\left({\Delta d_{\rm Moon}\over d_{\rm Moon}}\right) \sim 
\left({(\Delta \phi)_{\rm GW}\over \phi_\Earth}\right)\times
{1\over 2} \left({v_{\rm Moon}t_{\rm cross}\over d_{\rm
    Moon}}\right)^2 \sim 0.3 \times 10^{-11}\left[{(\Delta M)_{\rm
      GW}\over M_{\rm BH}}\right] ~,
\label{four}
\end{equation}
where $v_{\rm Moon}\approx 1~{\rm km~s^{-1}}$ is the Moon's orbital
speed, and the geometric calculation ignored the small eccentricity in
the Moon's orbit.

The above increase in distance as a result of the motion of the Moon
relative to Earth is of the same magnitude as the known ``memory
effect''
\citep{1974SvA....18...17Z,1985ZhETF..89..744B,1991PhRvL..67.1486C,2012CQGra..29u5003B},
for which the permanent change in separation between free-floating
objects of negligible mass initially at rest relative to each other,
is also of order, $(\Delta d_{\rm Moon}/d_{\rm Moon})\sim [(\Delta
  \phi)_{\rm GW}/c^2]$.

The resulting permanent change $\Delta d_{\rm Moon}\sim 1~{\rm
  mm}[(\Delta M)_{\rm GW}/ M_{\rm BH}]$ is above the ultimate
sensitivity threshold of lunar ranging \citep{2012CQGra..29r4005M},
$(\Delta d_{\rm Moon}/d_{\rm Moon})\sim 10^{-14}$, and could be
measured for future merger events. 

A tight binary of black holes with individual masses $\sim 2\times
10^6M_\odot$ and a separation $a$ would merge on a timescale of $\sim
40~{\rm yr}~(a/10^{14}~{\rm cm})^4$ \citep{1964PhRv..136.1224P}. The
existence of such a binary is not ruled out by the orbits of the
S-stars which are observed at much larger distances, $\gtrsim
10^{15}~{\rm cm}$ \citep{2010MNRAS.409.1146G}.

The permanent displacement from the memory effect would increase
slightly the eccentricity of stellar binaries at wide separations
$\gtrsim 10^{16}~{\rm cm}$, but this imprint is not detectable at the
precision enabled by astronomical surveys such as Gaia
\citep{2022MNRAS.512.3383H}, even when considering the increase in its
amplitude with decreasing Galactocentric distance.

\section{Limiting Flux Versus Redshift as a Flag of New Physics}

According to General Relativity, the maximum possible luminosity of a
source which is bound by its own gravity, equal to its total
rest-mass energy, $Mc^2$, divided by the light crossing-time of its
gravitational radius, $GM/c^2$. This limit applies to all possible
carriers of energy, including gravitational waves, elementary
particles such as neutrinos, or electromagnetic radiation. Packing the
energy to a smaller scale would result in an implosion to a black hole
according to the hoop conjecture \citep{2021arXiv210107353P}, and a
shorter emission time would require faster than light travel
\citep{2021PhRvD.104l4079S,2021IJMPD..3042026J,2018PhRvD..97h4013C,1999astro.ph.12110H}. These
considerations are purely classical and do not involve quantum
mechanics.

The ratio between the maximum emission energy and the minimum emission
time is independent of mass $M$, implying that the maximum luminosity
is a universal constant, combining the speed of light $c$ and
Newton's constant $G$,
\begin{equation}
L_{\rm max}={c^5\over G}= 3.64\times 10^{59}~{\rm erg~s^{-1}} .
\label{onet}
\end{equation}

A similar argument can be applied to any self-gravitating system of
size $r$, where the characteristic velocity $v$ is dictated by the
virial theorem, $v^2\sim GM/r$, and the characteristic energy release time
is limited by the crossing-time, $\sim r/v$, so that the limiting
output power is $\sim v^5/G$, smaller by a factor of $\sim (v/c)^5$
than the universal limit in equation~(\ref{onet}).

Given the luminosity distance as a function of redshift, $d_L(z)$, the
above limit gives a maximum energy flux that can be observed from a
cosmological source which emits isotropically any form of radiation or
relativistic particles,
\begin{equation}
f_{\rm max}(z)={L_{\rm max}\over 4\pi d_L^2(z)} .
\label{twot}
\end{equation}

To quantify this universal flux limit, I use an analytic approximation
to $d_L(z)$ which is accurate to a sub-percent level
\citep{2012PThPh.127..145A} for the standard flat cosmology with a
matter density parameter $\Omega_m=0.32$ and a Hubble constant of
$70~{\rm km~s^{-1}~Mpc^{-1}}$ \citep{2020A&A...641A...6P}. This gives,
\begin{equation}
f_{\rm max}(z)={13.26~{\rm erg~s^{-1}~cm^{-2}}\over 
(1+z)^2\left\{\phi(2.13) -(1+z)^{-1/2}\phi[2.13(1+z)^{-3}]\right\}^2} ,
\label{threet}
\end{equation}
where,
\begin{equation}
\phi(x) \equiv {1+1.32x+0.4415x^2+0.02656x^3\over 1+1.392x+0.5121x^2+0.03944x^3},
\label{fourt}
\end{equation}
and $\phi(2.13)=0.91$.

In the high-redshift limit, $z\gg 1$, I get the simple result,
\begin{equation}
f_{\rm max}(z)={15.93~{\rm erg~s^{-1}~cm^{-2}}\over 
(1+z)^2[1-\sqrt{1.21/(1+z)}]^2} .
\label{fivet}
\end{equation}
This flux limit can be used to set an upper limit on the redshift of a
source with an unknown origin.

For comparison, a flux of $\sim 15~{\rm erg~s^{-1}~cm^{-2}}$ is
generated by local blackbody radiation at a temperature of
23~K, about ten times hotter than the cosmic microwave background
today.

The highest luminosities for astrophysical sources are expected to
occur during the formation of compact objects, in the form of
gravitational waves or a $\gamma$-ray burst for a black hole or
neutrinos for a neutron star. A violation of the limit in
equation(\ref{fourt}) for an isotropic, self-gravitating source with a
known redshift $z$, would flag new physics.

The redshift of a cosmological source can be inferred from the
spectral lines of its host galaxy or the Ly$\alpha$ absorption imprinted 
by the intergalactic medium.

The limit in equation~(\ref{fourt}) should be multiplied by a
correction factor, $f_\Omega=(4\pi/\Delta \Omega)$, for a source
radiating its energy into a limited solid angle $\Delta \Omega$.

\section{Implications of Modified Inertia at Low Accelerations for Rockets}

MOdified Newtonian Dynamics (MOND) was proposed by Milgrom four
decades ago \citep{1983ApJ...270..365M} to explain the flat rotation
curves of galaxies and the baryonic Tully-Fisher relation
\citep{2020SHPMP..71..170M,2012AJ....143...40M,2021AJ....162..202M}. An
attractive interpretation of MOND is that at accelerations of
magnitude, $a\ll a_0=1.2\times 10^{-8}~{\rm cm^{-2}}$, the inertia of
an object of mass $m$ satisfies a modified equation of motion in
response to a force $F$
\citep{2011arXiv1111.1611M,2015CaJPh..93..107M},
\begin{equation}
m{a^2\over a_0}=F .
\label{oneth}
\end{equation}
Below I consider the implications of modified inertia for a rocket
whose fuel burns so as to produce a steady low-acceleration,
$a\ll a_0$. 

The force (momentum delivered per unit time) acting on a
rocket, is given by the mass ablation rate, $\dot{m}$, times the
exhaust speed of the ablated gas relative to the rocket, $v_{\rm exh}$,
\begin{equation}
F= -\dot{m}v_{\rm exh} .
\label{twoth}
\end{equation}
For a constant acceleration, the solution to equations (\ref{oneth}) and
(\ref{twoth}) is,
\begin{equation}
\left({m_{\rm initial}\over m_{\rm final}}\right)=
\exp\left\{\left({a\over a_0}\right)\left({{v_{\rm final}-v_{\rm
      initial}\over v_{\rm exh}}}\right)\right\} ,
\label{threeth}
\end{equation}
where the subscripts `initial' and `final' refer to the initial and
final values of the rocket mass and speed.  This result differs from
the standard Tsiolkovsky solution to the rocket equation
\citep{2000eaa..bookE4067} by the suppression factor $({a/a_0})$ in
the exponent. Whereas the amount of fuel that needs to be carried
grows exponentially with terminal speed in the standard Tsiolkovsky
solution, a modified inertia offers the prospects of reaching high
speeds by carrying much less fuel. This allows for intergalactic
travel at a modest fuel-to-payload mass ratio.

As a concrete example, consider an intergalactic journey at a final
speed of $v_{\rm final}\sim 300~{\rm km~s^{-1}}$, an order of
magnitude faster than the rockets launched so far by humans. For
standard chemical fuel, this terminal speed exceeds the exhaust speed
by a factor $({v_{\rm final}/v_{\rm exh}})\sim 10^2$
\citep{2004cdip.book.....G}. Thus, in order to achieve this terminal
speed through an average acceleration magnitude $(a/a_0)\sim 0.01$ in
free space, the required fuel mass would be comparable to the payload
mass, $(m_{\rm initial}-m_{\rm final})\sim 1.7m_{\rm final}$. At this
acceleration, the above terminal speed is obtained over a timescale,
$t\sim 8 {\rm Gyr}$, comparable to the remaining lifetime of the
Sun. During this time, the rocket would be able to traverse a distance
${1\over 2}a t^2\sim 1.2~{\rm Mpc}$, all the way to the edge of the
Local Group of galaxies. Of course, additional fuel would be needed to
overcome the binding energy of the Earth, the Sun and the Milky-Way
galaxy.
 
The validity of the modified rocket equation can be tested by
launching our own low-acceleration rocket or by finding low-acceleration
rockets which arrived to our vicinity from great distances. It is
unclear which approach is more likely to bear fruit as the first
direct test of the modified inertia interpretation of MOND.

While escaping from the Earth, the Sun and the local Galactic
environment, a rocket would need to overcome gravitational
accelerations in excess of $a_0$. However, it is the net acceleration
that counts in MOND, and so the rocket engine can be designed to
produce just a little above what is needed to escape and stay in the
MOND acceleration regime.  This requires fine tuning of the rocket
thrust, but is possible.

\section{The CMB Dipole in MOND}

The dipole anisotropy of the Cosmic Microwave Background (CMB)
corresponds to a velocity of $\sim 300~{\rm
  km~s^{-1}}=10^{-3}c$~\citep{2021PhRvL.127j1301F}. It is commonly
assumed that this velocity is induced by gravity across cosmic scales,
as a result of an average cosmic acceleration of, $a\sim 10^{-3}
H_0c$, over the age of the Universe $\sim H_0^{-1}$, where $H_0\sim
70~{\rm km~s^{-1}~Mpc^{-1}}$ is the present-day Hubble constant
\citep{2021ApJ...919...16F}.

This acceleration $a$ of our cosmic neighborhood is by lower by a
factor of $\sim 5\times 10^{-3}$ than the threshold acceleration,
$a_0=1.2\times 10^{-8}~{\rm cm~s^{-2}}\sim 0.2 H_0c$, in Modified
Newtonian Dynamics (MOND)~\citep{2020SHPMP..71..170M}. Hence, in MOND
- the gravitational acceleration $g$ required to induce the CMB dipole
is,
\begin{equation}
g\sim \left( {a^2\over a_0}\right)\sim 2.5 \times 10^{-5} H_0c .
\label{one}
\end{equation}

Interestingly, this gravitational acceleration could result from a
primordial amplitude of density perturbations on the scale of the
cosmic horizon, $\delta \sim (g/H_0c) \sim 2.5 \times 10^{-5}$,
comparable to the amplitude of primordial fluctuations on horizon
scales~\citep{1993ppc..book.....P}.

A nearby perturbation of a larger amplitude, like a supercluster,
would induce within MOND a much larger CMB dipole than observed.
Local galaxy survey do not converge as of yet to the observed diopole
direction and
amplitude~\citep{2008MNRAS.386.2221L,2018JCAP...01..013M}.

\section{Quantum Tunneling of Fuzzy Dark Matter Out of Satellite Galaxies}

The axion explains the lack of observed CP violation in quantum
chromodynamics and is also an attractive dark matter candidate
\citep{1977PhRvL..38.1440P,1978PhRvL..40..223W,1978PhRvL..40..279W}. Ultra-light
axions (ULAs) with masses $\sim 10^{-22}~{\rm eV}$, also known as
fuzzy dark matter
\citep{2000PhRvL..85.1158H,2017PhRvD..95d3541H,2019PhRvD..99f3517V,2021PhRvL.126g1302R},
are of particular interest in relieving small-scale challenges to the
cold dark matter paradigm, stemming from discrepancies between
observations and simulations of galaxies \citep{2017ARA&A..55..343B}.

Dwarf galaxies (DG) in the halo of the Milky-Way galaxy are subject to
stripping by the gravitational tidal force
\citep{2018MNRAS.476.3816F,2016ApJ...819...20G,2020A&A...638A.133L,2022MNRAS.510.2186G,2022MNRAS.511.6001E}. To
leading order, the tidal force scales in proportion to the distance
from the DG center, $r$, implying a tidal potential that scales as
$r^2$. This modifies the binding gravitational potential of an
isolated DG, $\phi(r)$. Along the axis connecting the two galaxies, it
yields a net potential,
\begin{equation}
V(r) = m_a\left[\phi(r)-\sigma^2\left({r\over r_t}\right)^2\right],
\label{oned}
\end{equation}
where $\sigma$ is the characteristic 3D velocity dispersion of the DG
and $r_t$ is the tidal radius. Classically, dark matter particles at
$r>r_t$ are stripped from the DG. But quantum-mechnical tunneling
allows particles which are classically-bound to leak through the
gravitational potential barrier and escape from $r<r_t$ as a result of
the extended tail of their wavefunction. This constitutes a new source
for evaporation of dark matter particles with a small mass, such as
ULAs.

The characteristic escape timescale of ULAs from $V(r)$ due to
quantum-mechanical tunneling is,
\begin{equation}
T_{\rm esc}\sim \left({r_t\over \sigma}\right)\exp\left\{-{r_t\over \lambda_a}\right\} ,
\label{twod}
\end{equation}
and
\begin{equation}
\lambda_a= {\hbar\over 2 \sqrt{m_a^2\vert\langle\phi\rangle \vert}},
\label{threed}
\end{equation}
with the virial theorem implying that the average kinetic energy of
the bound particles is roughly half of their average gravitational
binding energy, namely $\langle \phi\rangle \sim - \sigma^2$, and so
\begin{equation}
\lambda_a= \left[{1~{\rm kpc}\over (m_a/10^{-22}~{\rm
  eV})(\sigma/10~{\rm km~s^{-1}})}\right] .
\label{fourd}
\end{equation}

Given that some dark-matter-rich DGs in the Milky-Way halo have
$r_t\lesssim 0.5~{\rm kpc}$, $\sigma\lesssim 5~{\rm km~s^{-1}}$
\citep{2011ASSP...27..229L,2016ApJ...819...20G,2020A&A...638A.133L,2022MNRAS.510.2186G,2022MNRAS.511.6001E},
and $(r_t/\sigma)\lesssim 10^8~{\rm yr}$, their possession of dark
matter over a Hubble time implies $T_{\rm esc}\gtrsim 10^{10}~{\rm
  yr}$ and therefore $m_a\gtrsim 2\times 10^{-21}~{\rm eV}$, based on
equations (\ref{one}-\ref{four}). 

This new constraint based on inevitable escape resulting from
quantum-mechanical tunneling, rules out the preferred mass range for
ULAs as dark matter.

\section{Gravitational Wave Acceleration to Relativistic Energies}

Consider a relativistic charged particle with a mass $m$, a charge
$q$, a velocity vector ${\bf v}={\bf \beta}c$ and a Lorentz factor
$\gamma=(1-\beta^2)^{-1/2}$, that gyrates at the cyclotron frequency,
$\Omega_c=(qB/mc)$ in a uniform magnetic field along the $z$-axis,
${\bf B}=B{\hat z}$.  If the cyclotron frequency resonates with the
Doppler shifted frequency of a wave propagating along the magnetic
field direction, the particle would witness steady acceleration at a
fixed phase of the wave crest.

For a transverse wave propagating at the speed of light, $c$, the
Doppler-shifted cyclotron resonance is given by,
\begin{equation}
\Omega_c=D\omega,
\label{zero}
\end{equation}
where $D\equiv \gamma(1-\beta_z)$ is the Doppler factor along the
$z$-axis~\citep{1986rpa..book.....R}, and $\omega$ is the wave
frequency in the background frame of reference.

The relativistic equation of motion of the charged particle in the
presence of either an electromagnetic
wave~\citep{1986PhRvA..33.1828L,1987PhRvA..35.1692L} or a
gravitational wave~\citep{2001PhRvD..64b4013S} propagating along the
$z$-axis, admits the same identity,
\begin{equation}
{d\gamma\over dt}={d(\gamma \beta_z)\over dt}.
\label{one}
\end{equation}
This implies the remarkable result that the Doppler factor
$D=\gamma(1-\beta_z)$ is a constant of motion, guaranteeing that the
resonance condition in equation~(\ref{zero}) will be maintained at all
times if it is satisfied initially, even as the particle gains energy.

Under resonance, the Lorentz factor of a relativistic particle with
$\gamma \gg 1$ grows steadily over time,
\begin{equation}
{d\gamma\over dt}=\Omega_c\alpha ,
\label{two}
\end{equation} 
where for an electromagnetic wave~\citep{1986PhRvA..33.1828L}:
\begin{equation}
\alpha_{\rm EM}=\sqrt{{2\over
    \gamma}}\times \left({qA\over mc}\right),
\label{three}
\end{equation}
with ${\bf A}$ being the vector potential. For a
gravitational wave~\citep{2001PhRvD..64b4013S},
\begin{equation}
\alpha_{\rm GW}=2h,
\label{four}
\end{equation}
with $h$ being the dimensionless wave amplitude.

The electromagnetic wave case serves as the basis for a novel
acceleration scheme, the so-called ``autoresenance laser
accelerator''~\citep{1986PhRvA..33.1828L}. Quasi-neutrality is
perfectly maintained in a symmetric electron-positron
plasma~\citep{1987PhRvA..35.1692L}.

Equations~(\ref{two}) and (\ref{four}) imply that a plane-parallel
gravitational wave of a constant amplitude $h$ propagating along a
constant magnetic field ${\bf B}$, would accelerate charged particles
at a constant rate to arbitrarily high energies.

However, for astrophysical sources of gravitational waves, the wave
amplitude $h$ declines inversely with distance from the source. Given
a magnetic field $B$ oriented radially from the source over a
coherence length $\ell$, equation~(\ref{two}) implies that the
particle's Lorentz factor would reach a maximum value of,
\begin{equation}
\gamma_{\rm max}\sim 1+ 2h \Omega_c\left({\ell\over c}\right) .
\end{equation} 
Since $2h\lesssim (R_{\rm Sch}/\ell)$~\citep{1986bhwd.book.....S}, we get
\begin{equation}
\gamma_{\rm max}\lesssim 1+ \Omega_c t_{\rm Sch},
\end{equation}
where $t_{\rm Sch}=(R_{\rm Sch}/c)=10^{-4}~{\rm s}\times(M/10M_\odot)$ is
the light crossing-time for the Schwarzschild radius $R_{\rm
  Sch}=(2GM/c^2)= 30~{\rm km}\times(M/10M_\odot)$, of a gravitational wave
source of total mass $M$.

The cyclotron frequency for electrons is $\Omega_{c,e}=180~{\rm
  Hz}\times(B/10\mu{\rm G})$ (whereas for protons, $\Omega_c$ is smaller by
the particle mass-ratio of $1.836\times 10^3$), can resonate with the
frequency of gravitational waves generated in the final coalescence
phase of binaries composed of stellar-mass black holes or neutron
stars, which produce $\omega \sim
10$--$10^4$~Hz~\citep{2022Galax..10...36C,2022Galax..10...91V}.

Under favorable conditions, the cyclotron resonance could boost the
energies of relativistic electrons or protons in the plasma
surrounding compact gravitational-wave sources. If the magnetic field
originates from the source - as expected in the case of neutron star
mergers, its dipole amplitude would decline inversely with distance
cubed, $B\propto \ell^{-3}$, out to the scale where the interstellar
magnetic field will dominate.

The autoresonant acceleration by gravitational waves can heat
relativistic electrons or protons in the vicinity of mergers of
compact objects. This, in turn, could trigger synchrotron emission by
the accelerated electrons that would result in an electromagnetic
counterpart to the gravitational wave signal.

In principle, the cyclotron autoresonance could potentially lead to
electromagnetic afterglows of the type reported for the black hole
merger GW150914~\citep{2016ApJ...819L..21L,2018PhRvD..97h3008D} or the
neutron star merger
GW170817/GRB170817A~\citep{2022Natur.610..273M}. Detailed modeling is
needed for the expected electromagnetic counterpart in specific
environments.

\section{The Horizon of Intergalactic Travel}

The distant future of the standard LCDM cosmological model implies
that all galaxies beyond the Milky-Way will eventually exit from our
event
horizon~\citep{2000ApJ...531...22K,2002PhRvD..65d7301L,2003NewA....8..439N,2003ApJ...596..713B,2004NewA....9..573N,2005PhRvD..72j7302H,2011JCAP...04..023L}. 

The event horizon is conventionally defined based on the speed of
light. The limiting distance shrinks to much smaller scales
for non-relativistic motions. Here, we consider the limit imposed by
the future exponential expansion of an LCDM cosmology on the maximum
comoving distance that can be traversed by non-relativistic
astrophysical objects, such as hypervelocity stars or black holes.

The mass budget of the standard LCDM model is currently dominated by a
cosmological constant with a density parameter, $\Omega_\Lambda
\approx 0.7$~\citep{2020A&A...641A...6P}. The future evolution of the
scale factor follows the Friedmann-Lemaitre-Robertson-Walker equation
for a flat geometry,
\begin{equation}
\left({\dot a}\over a\right)^2=H_0^2\left[\Omega_\Lambda
  +{(1-\Omega_\Lambda)\over a^3}\right],
\label{one}
\end{equation}
where $H_0\approx 70~{\rm km~s^{-1}~Mpc^{-1}}$ is the current Hubble
constant.  The evolution of the scale factor in the distant future,
$t\gg t_0$, is to a good approximation, $a\approx
\exp\{\sqrt{\Omega_\Lambda}H_0(t-t_0)\}$, where $t_0=13.8~{\rm Gyr}$
is the present cosmic time.

An object moving through intergalactic space at a non-relativistic
peculiar velocity, $v$, traverses an infinitesimal physical distance
$adr$ over a time interval, $dt$, where $r$ is the radial comoving
coordinate. Since the object's momentum declines inversely with the scale
factor (equivalent to redshifting its de Broglie wavelength), we get,
\begin{equation}
v={adr\over dt}={v_0\over a},
\label{two}
\end{equation}
where $v_0$ is the current value of the peculiar velocity.

The time-integration of equation~(\ref{two}) yields,
\begin{equation}
\Delta r= (r-r_0)=v_0\int_{t_0}^{t}{dt\over a^2} \approx r_{\rm
  max}\left[1-e^{-2\sqrt{\Omega_\Lambda}H_0t}\right] ,
\label{three}
\end{equation}
implying that future intergalactic travel in LCDM is limited to
traversing a maximum comoving distance, even for an infinite travel 
time,
\begin{equation}
r_{\rm max}= {v_0\over 2\sqrt{\Omega_\Lambda}H_0} = 26~{\rm Mpc} \times \left({v_0
\over 3\times 10^3~{\rm km~s^{-1}}}\right) .
\label{four}
\end{equation}

Equation (\ref{four}) sets an upper limit to the distance that
galactic winds, driven by present-day supernovae or supermassive black
holes~\citep{2018MNRAS.473.4077P}, can reach and enrich the
intergalactic medium (whose mass can slow them down further).

Hypervelocity stars~\citep{2015ARA&A..53...15B}, which exit the halo
of the Milky Way galaxy at a speed below a few thousand ${\rm
  km~s^{-1}}$, will not go beyond a comoving distance of a few tens of
Mpc even after trillions of years, the lifetime of low-mass
stars~\citep{2005AN....326..913A}. The gravitational-wave recoil of
the black hole remnant from a merger of black hole binaries is also
limited to similar
speeds~\citep{2007ApJ...662L..63S,2007PhRvL..99d1103L}. 

However, the ejection speed could be larger for triple systems
involving three black holes~\citep{2012MNRAS.422.1306K} or stars
interacting with a black hole binary~\citep{2015ApJ...806..124G}. The
traversed distance will be largest for semi-relativistic stars,
ejected by binaries of supermassive black
holes~\citep{2016AMSA....1..183L}. For $v_0\sim 0.1c$, these stars can
traverse 260~comoving Mpc.

To reach beyond the Virgo cluster of galaxies, requires an initial
peculiar speed over $0.01c=3\times 10^3~{\rm km~s^{-1}}$, a hundred
times faster than the chemical rockets that humanity launched to space
so far.

\bigskip
\bigskip
\section*{Acknowledgements}

This work was supported in part by a grant from the Breakthrough Prize
Foundation and by Harvard's {\it Black Hole Initiative} which is
funded by grants from GBMF and JTF. I thank Greg Laughlin, Fabio
Pacucci and Sunny Vagnozzi for helpful comments on the final
manuscript.

\bigskip
\bigskip
\bigskip

\bibliographystyle{aasjournal}
\bibliography{c}
\label{lastpage}
\end{document}